\documentstyle[prl,aps,psfig]{revtex}
\newcommand{\beq}{\begin{equation}}
\newcommand{\eeq}{\end{equation}} 
\newcommand{\beqa}{\begin{eqnarray}}
\newcommand{\eeqa}{\end{eqnarray}} 
\newcommand{\ba}{\begin{array}} 
\newcommand{\ea}{\end{array}} 

\begin{document}
\draft

\twocolumn[\hsize\textwidth\columnwidth\hsize\csname
@twocolumnfalse\endcsname

\widetext 
\title{Effective wave-equations for the dynamics \\
of cigar-shaped and disc-shaped Bose condensates} 
\author{L. Salasnich$^1$, A. Parola$^2$ and L. Reatto$^1$} 
\address{$^1$Istituto Nazionale per la Fisica della Materia, 
Unit\`a di Milano Universit\`a, \\ 
Dipartimento di Fisica, Universit\`a di Milano, \\
Via Celoria 16, 20133 Milano, Italy \\
$^2$Istituto Nazionale per la Fisica della Materia, 
Unit\`a di Como, \\
Dipartimento di Scienze Fisiche, Universit\`a dell'Insubria, \\
Via Valleggio 11, 23100 Como, Italy}

\maketitle 

\begin{abstract}
Starting from the 3D Gross-Pitaevskii equation and 
using a variational approach, we derive an effective 
1D wave-equation that describes the axial 
dynamics of a Bose condensate confined in an external 
potential with cylindrical 
symmetry. The trapping potential is harmonic in the transverse 
direction and generic in the axial one. 
Our equation, that is a time-dependent 
non-polynomial nonlinear Schr\"odinger equation (1D NPSE), 
can be used to model cigar-shaped condensates, whose dynamics is 
essentially 1D. We show that 1D NPSE gives much more accurate results than 
all other effective equations recently proposed. 
By using 1D NPSE we find analytical solutions 
for bright and dark solitons, which generalize the 
ones known in the literature. 
We deduce also an effective 2D non-polynomial 
Schr\"odinger equation (2D NPSE) that models disc-shaped 
Bose condensates confined in an external trap that is harmonic 
along the axial direction and generic in the transverse direction. 
In the limiting cases of weak and strong interaction, 
our approach gives rise to Schr\"odinger-like equations 
with different polynomial nonlinearities. 
\\
PACS Numbers: 32.80.Pj; 67.65.+z; 51.10.+y 
\\
\end{abstract} 

]

\narrowtext 

\vskip 0.5cm 

\section{Introduction} 

Bose condensates are nowadays routinely produced by  
many experimental groups all over the world. 
The thermal and dynamical properties of Bose condensates have been 
investigated with different atomic species and various 
trap geometries. An interesting theoretical problem 
is the derivation of 1D and 2D equations, describing 
cigar-shaped and disc-shaped condensates, 
respectively. At zero temperature, a good theoretical 
tool for the study of the dynamics of dilute condensates is 
the time-dependent 3D Gross-Pitaevskii equation \cite{1}. 
In the case of reduced dimensionality, various approaches 
have been adopted to derive effective equations from the 
3D Gross-Pitaevskii equation \cite{2,3,4}. 
\par 
In this paper we analyze both cigar-shaped and disc-shaped 
condensates. By using a variational approach, we obtain an effective 1D 
time-dependent non-polynomial nonlinear Schr\"odinger equation 
that describes the axial dynamics of a Bose condensate 
confined in an external potential with cylindrical 
symmetry. We demonstrate that our equation exactly 
reproduces previous findings in the limits of weak-coupling 
and strong-coupling. Moreover, we show that our variational 
approach is more accurate than all other recently 
proposed procedures in the evaluation of both static 
and dynamical properties of the condensate. 
We also investigate the 2D reduction of the 3D Gross-Pitaevskii equation. 
In this case it is not possible to analytically determine 
an single effective 2D wave-equation which describes the dynamics 
of disc-shaped condensates. Nevertheless, analytical 
equations can be found in the weakly-interacting limit 
and in the strongly-interacting limit. 

\section{Effective 1D equation} 

The 3D Gross-Pitaevskii equation (3D GPE), which describes 
the macroscopic wavefunction $\psi({\bf r},t)$ of the 
Bose condensate, is given by 
\beq 
i\hbar {\partial \over \partial t}\psi ({\bf r},t)= 
\left[ -{\hbar^2\over 2m} \nabla^2 
+ U({\bf r}) + g N |\psi ({\bf r},t)|^2 \right] \psi({\bf r},t)  \; , 
\eeq  
where $U({\bf r})$ is the external trapping potential and 
$g={4\pi \hbar^2 a_s/m}$ is the scattering amplitude and $a_s$ 
the s-wave scattering length \cite{1}. $N$ is the number of condensed 
bosons and the wave-function is normalized to one. 
Note that the 3D GPE is accurate to describe a condensate 
of dilute bosons only near zero temperature, where thermal 
excitations can be neglected \cite{5}. 
\par 
The 3D GPE can be obtained by using the quantum least action 
principle, i.e. 3D GPE is the Euler-Lagrange equation 
of the following action functional 
$$
S = \int dt d{\bf r} \; 
\psi^*({\bf r},t) \left[ i\hbar {\partial \over \partial t} 
+ {\hbar^2 \over 2 m} \nabla^2 - U({\bf r}) 
\right. 
$$
\beq
\left. 
-{1\over 2}gN |\psi ({\bf r},t)|^2 \right] \psi({\bf r},t)  \; . 
\eeq 
We consider an external potential with cylindrical symmetry. 
In particular we analyze a trapping potential 
that is harmonic in the transverse direction 
and generic in the axial direction: 
$U({\bf r})={1\over 2}m\omega_{\bot}^2(x^2+y^2) + V(z)$. 
We want to minimize the action functional $S$ 
by choosing an appropriate trial wavefunction.  
A natural choice \cite{3} is the following 
\beq 
\psi({\bf r},t) = \phi(x,y,t;\sigma(z,t)) \; f(z,t) \; , 
\eeq 
where both $\phi$ and $f$ are normalized and $\phi$ is represented 
by a Gaussian:  
\beq
\phi(x,y,t;\sigma(z,t)) = { e^{-(x^2+y^2)\over 2 \sigma(z,t)^2} 
\over \pi^{1/2} \sigma(z,t)} \; . 
\eeq 
The variational functions $\sigma(z,t)$ and $f(z,t)$ 
will be determined by minimizing the action functional 
after integration in the $(x,y)$ plane. 
The choice of a Gaussian shape for the condensate 
in the transverse direction is 
well justified in the limit of weak interatomic coupling, 
because the exact ground-state of the linear Schr\"odinger 
equation with harmonic potential is a Gaussian. Moreover, 
for the description of the collective
dynamics of Bose-Einstein condensates, 
it has already been shown that the variational technique based 
on Gaussian trial functions leads to 
reliable results even in the large condensate number limit \cite{6,7}. 
\par 
We assume that the transverse wavefunction $\phi$ is slowly 
varying along the axial direction with respect 
to the transverse direction \cite{3}, 
i.e. $\nabla^2 \phi \simeq \nabla_{\bot}^2 \phi$  
where $\nabla_{\bot}^2={\partial^2 \over \partial x^2}+  
{\partial^2 \over \partial y^2}$.  
By inserting the trial wave-function in (2)  
and after spatial integration along $x$ and $y$ variables  
the action functional becomes  
$$ 
S = \int dt dz \;  f^* \left[ i\hbar 
{\partial \over \partial t}  + {\hbar^2\over 2 m} 
{\partial^2\over \partial z^2} - V -  {1\over 2} 
g N {\sigma^{-2}\over 2\pi} |f|^2 \right.  
$$ 
\beq 
\left.  - {\hbar^2 \over 2m}\sigma^{-2} - 
{m\omega_{\bot}^2\over 2} \sigma^2   \right] f \; .  
\eeq 
The Euler-Lagrange equations with respect  
to $f^*$ and $\sigma$ read  
$$  
i\hbar {\partial \over \partial t}f=  
\left[ -{\hbar^2\over 2m} {\partial^2\over \partial z^2}  
+ V + g N {\sigma^{-2}\over 2\pi} |f|^2 \right.  
$$ 
\beq 
\left.  + \left({\hbar^2 \over 2m}\sigma^{-2} 
+ {m\omega_{\bot}^2\over 2} \sigma^2  \right) \right] f  \; ,  
\eeq   
\beq {\hbar^2 \over 2m}\sigma^{-3} - 
{1\over 2}m\omega_{\bot}^2 \sigma  + 
{1\over 2} g N {\sigma^{-3}\over 2\pi} |f|^2 = 0 \; .  
\eeq 
The second Euler-Lagrange equation reduces to an algebric  
relation providing a one to one correspondence between $\sigma$  
and $f$: $\sigma^2 = a_{\bot}^2 \sqrt{1 + 2 a_s N |f|^2}$,  
where $a_{\bot}=\sqrt{\hbar \over m\omega_{\bot}}$ 
is the oscillator  length in the transverse direction. 
One sees that $\sigma$  depends implicitly on $z$ and $t$ 
because of the  space and time dependence of $|f|^2$.  
Inserting this result in the first equation one finally obtains  
$$  
i\hbar {\partial \over \partial t}f=  
\left[ -{\hbar^2\over 2m} {\partial^2\over \partial z^2}  
+ V + {g N \over 2\pi a_{\bot}^2} {|f|^2\over 
\sqrt{1+ 2a_sN|f|^2} }  \right.  
$$ 
\beq 
\left.  
+ {\hbar \omega_{\bot}\over 2}   
\left( {1\over \sqrt{1+ 2 a_sN|f|^2} } 
+ \sqrt{1+ 2a_sN|f|^2} \right) \right] f  \; . 
\eeq  
This equation is the main result of our paper. 
It is a time-dependent non-polynomial nonlinear 
Schrodinger equation (1D NPSE). 
\par
We observe that from 1D NPSE in certain limiting cases 
one recovers familiar results. 
In the weakly-interacting limit $a_sN|f|^2 <<1$ 
one has $\sigma^2=a_{\bot}^2$ and the previous equation reduces to 
\beq 
i\hbar {\partial \over \partial t}f= 
\left[ -{\hbar^2\over 2m} {\partial^2\over \partial z^2} 
+ V + {g N \over 2\pi a_{\bot}^2} |f|^2 \right] f  \; ,  
\eeq 
where the additive constant $\hbar \omega_{\bot}$ has been omitted because 
it does not affect the dynamics. This equation is a 
1D Gross-Pitaevskii equation. The nonlinear coefficient $g'$ 
of this 1D GPE can be thus obtained from the nonlinear coefficient 
$g$ of the 3D GPE by setting $g'=g/(2\pi a_{\bot}^2)$. 
This ansatz has been already used by various authors, 
for example in Ref. \cite{2}. 
Note that the limit $a_sN|f|^2<<1$ is precisely 
the regime where the healing length is larger than $\sigma$. 
In this regime the cigar-shaped condensate is quasi-1D, 
as shown in a recent experiment \cite{8}. 
It is well known that in 1D and at ultra-low densities 
($a_sN|f|^2<< a_s^2/a_{\bot}^2<<1$) 
an interacting Bose gas becomes a 
Tonks gas, i.e. a gas of spinless Fermions \cite{9}. 
Such a transition cannot be described 
by the cubic 3D GPE equation and therefore by 
1D NPSE because in the Tonks regime the inter-atomic interaction 
cannot simply be approximated by a zero-range pseudo-potential 
in mean field approximation\cite{10}. 
Instead, in the strongly-interacting high density  
limit $a_sN|f|^2 >> 1$ (but $N|\psi|^2 a_s <<1$ to satisfy 
the diluteness condition) 
one finds $\sigma^2 = \sqrt{2} a_{\bot}^2 a_s^{1/2}N^{1/2} |f|$ and 
the 1D NPSE becomes 
\beq 
i\hbar {\partial \over \partial t}f= 
\left[ -{\hbar^2\over 2m} {\partial^2\over \partial z^2} 
+ V + {3\over 2}{g N^{1/2} 
\over 2\pi a_{\bot}^2 \sqrt{2a_s}} |f| \right] f  \; .   
\eeq 
In this limit, and in the stationary case, 
the kinetic term can be neglected (Thomas-Fermi approximation) 
and one finds the following analytical formula for 
the axial density profile 
\beq 
|f(z)|^2 = {2\over 9} {1\over (\hbar\omega_{\bot})^2 a_s N} 
\left( \mu' - V(z) \right)^2 \; , 
\eeq 
where $\mu'$ is the chemical potential, fixed by the 
normalization condition. 
It is important to stress that this 1D Thomas-Fermi density 
profile is quadratic in the term $\mu' -V(z)$. 
The same quadratic dependence is obtained starting 
from the Thomas-Fermi approximation of the 3D stationary GPE, 
i.e. neglecting the spatial derivatives in Eq. (1), 
and then integrating along $x$ and $y$ variables. 
In this way one finds a formula that differs from 
(11) only for the numerical factor which is $1/4$ instead of $2/9$. 
\par
To test the accuracy of the full 1D NPSE, Eq. (8), and compare it 
with other procedures proposed in the last few years, 
we numerically investigate the simple case of harmonic trapping  
also in the axial direction: $V(z)= {1\over 2} m \omega_z^2 z^2$. 
In this case, the 1D NPSE can be written in scaled units: 
$z$ in units of $a_z=\sqrt{\hbar\over m\omega_z}$, 
the oscillator length in the axial direction, and 
$t$ in units of $1/\omega_z$. 
In order to assess the accuracy of the various 1D approximations 
we have also solved numerically 
the 3D Gross-Pitaevskii equation (3D GPE), 
given by Eq. (1) with imaginary time \cite{11}. 
Note that the numerical solution of the 1D NPSE 
is not more time consuming than the solution of 
the standard 1D GPE. 
In Fig. 1 we plot the normalized density profile 
$\rho(z)=|f(z)|^2$ of the ground-state wave-function 
of a cigar-shaped condensate confined in a trap with an aspect ratio 
$a_z/a_{\bot}=10$. 

\begin{figure}
\centerline{\psfig{file=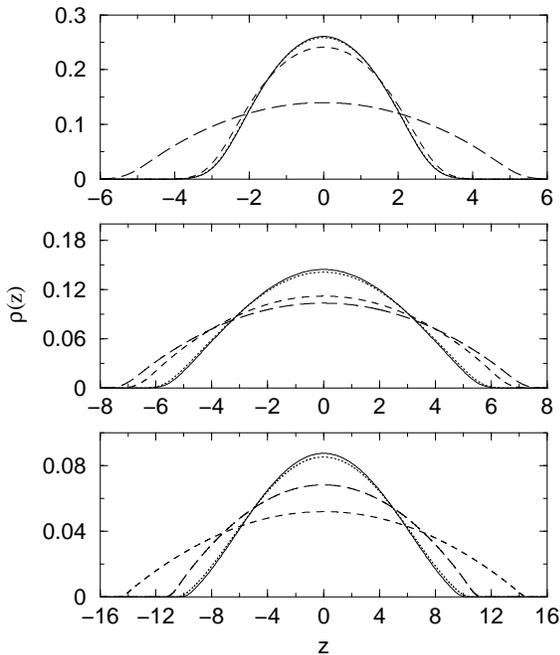,height=3.5in}}
\caption{Normalized density profile $\rho(z)=|f(z)|^2$ along the axial 
direction $z$ for the cigar shaped trap. 
Number of Bosons: $N=10^4$ and 
trap anisotropy: $\omega_{\bot}/\omega_z=10$. 
Four different procedures: 
3D GPE (solid line), 1D GPE (dashed line), CGPE 
(long-dashed line) and 1D NPSE (dotted line). 
From top to bottom: $a_s/a_z=10^{-4}$, 
$a_s/a_z=10^{-3}$, $a_s/a_z=10^{-2}$. 
Length $z$ in units of $a_z$ and density in units of $a_z^{-1}$.} 
\end{figure} 

We compare the results of four different procedures. 
The first procedure is the "exact" one, 
i.e. the solution of the 3D GPE. 
The second procedure is the numerical solution 
of the 1D Gross-Pitaevskii equation (1D GPE) given by Eq. (9) 
with an imaginary time. 
The third procedure is that proposed 
by Chiofalo and Tosi \cite{4}. 
In this case the nonlinear term of the 1D Gross-Pitaevskii equation 
is found by imposing that the 1D wave-function has the same chemical 
potential of the 3D one (CGPE). The fourth and last procedure 
is the numerical solution of our non-polynomial nonlinear Schr\"odinger 
equation (1D NPSE), i.e. Eq. (8) with imaginary time. 
As shown in Fig. 1, the 1D NPSE results are always very close 
to the "exact" ones and much better 
of the other approximations. Moreover, the CGPE procedure gives 
better results than the 1D GPE for large values of the 
scattering length where Eq. (9) is not reliable 
but in any case the 1D NPSE is superior. 
\par 
The very good performance of Eq. (8) is not limited 
to the ground-state. 
We have investigated the dynamics of the condensate 
by taking the previously calculated ground-state wavefunctions 
but changing the harmonic trap in the axial direction:  
from $V(z)={1\over 2}m\omega_z^2z^2$ to 
$V(z)={2\over 5} m\omega_z^2z^2$. In this way the 
condensate shows large collective shape oscillations along 
the $z$ axis. 

\begin{figure}
\centerline{\psfig{file=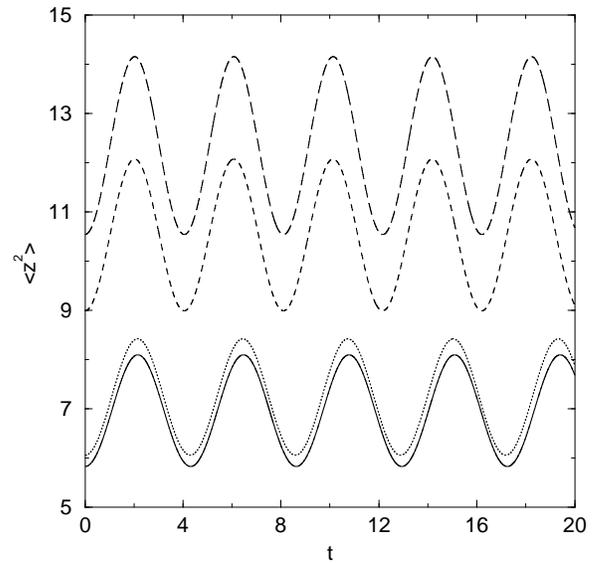,height=3.in}}
\caption{Squared amplitude $<z^2>$ as a function of time $t$. 
Number of Bosons: $N=10^4$ and 
trap anisotropy: $\omega_{\bot}/\omega_z=10$. 
Four different procedures: 
3D GPE (solid line), 1D GPE (dashed line), CGPE 
(long-dashed line) and 1D NPSE (dotted line). 
Scattering length: $a_s/a_z=10^{-3}$ 
Length $z$ in units of $a_z$, density in units of $a_z^{-1}$ 
and time $t$ in units of $1/\omega_z$.} 
\end{figure}

In Fig. 2 we plot the time evolution of the squared 
amplitude $<z^2>$ of the condensate in the axial direction, given by 
$<z^2> = \int dz z^2 \rho(z,t)$, where $\rho(z,t)=\int dx dy 
|\psi({\bf r},t)|^2$ in the case of the 3D GPE 
and $\rho(z,t)=|f(z,t)|^2$ in the other cases. 
Apart the better evaluation of the amplitude that is a consequence 
of the better evaluation of the ground-state wave-function, 
one sees that our 1D NPSE reproduces quite well 
also the "exact" sinusoidal behavior of the collective oscillation. 
For example, after four oscillations 
the relative error in the determination of the instant $t$ 
of minimum radius $<z^2>$ is about $1\%$. The other two procedures  
(CGPE and 1D GPE) clearly show a frequency delay 
with respect to the 3D GPE and 1D NPSE results: 
after four oscillations their relative errors,  
with respect to the 3D GPE result, 
are about $5\%$ and $6\%$ respectively. 
\par 
We remind that the 1D NPSE has been obtained 
by using a factorization of the 3D wavefunction 
with a variational ansatz for the transverse part of the wavefunction 
and neglecting the term ${\hbar^2 \over 2m}
\phi^* {\partial^2 \phi\over \partial z^2}$. 
Under the condition of the present computations 
the last assumption is fully justified because 
we have numerically verified that the ratio 
between the neglected term and the total energy 
ranges form $10^{-3}$ in the weak-coupling limit 
to $10^{-7}$ in the strong-coupling limit. 

\begin{figure}
\centerline{\psfig{file=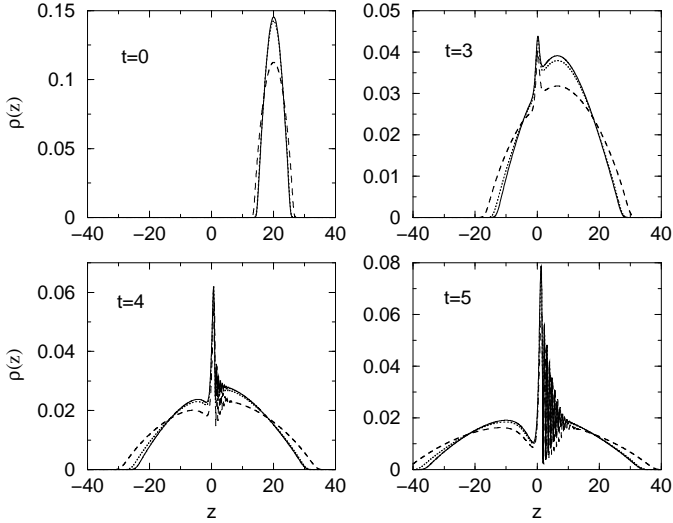,height=2.8in}}
\caption{Four frames of the axial density profile 
$\rho(z)$ of the Bose condensate tunneling through 
the Gaussian barrier (Eq. (3)). 
Start-up kinetic energy per particle 
of the condensate: $E_0=10$. 
Gaussian barrier parameters: $V_0=10$ and $\Sigma=1$. 
Interaction strength: $Na_s/a_z=10$. 
Comparison among 3D GPE (solid line), 1D NPSE (dotted line) 
and 1D GPE (dashed line). 
Length in units of $a_z=(\hbar /m\omega_z)^{1/2}$, 
density in units of $a_z^{-1}$, 
time in units of $\omega_z^{-1}$, and 
energy in units of $\hbar \omega_z$. } 
\end{figure}

\par 
To test the accuracy of 1D NPSE in the description 
of the dynamics of a cigar-shaped Bose condensate in 
more complex problem, we investigate the scattering and tunneling 
of the condensate on a Gaussian barrier. 
The initial wave function of the condensate is found 
by solving the equations with imaginary time 
and imposing a harmonic trapping potential 
also in the horizontal axial direction: 
\beq 
V(z)= {1\over 2}m\omega_z^2 (z-z_0)^2 \; . 
\eeq 
To have a cigar-shaped condensate 
we choose $\omega_{\bot}/\omega_z=10$. 
We set $z_0=20$, where $z_0$ is written 
in units of the harmonic length $a_z=(\hbar/m\omega_z)^{1/2}$. 
For $t>0$ the trap in the axial direction is switched off 
and a Gaussian energy barrier is inserted at $z=0$. The 
potential barrier is given by 
\beq
V(z)= V_0 \; e^{-z^2/ \Sigma^2} \; , 
\eeq 
where $V_0$ is the height of the potential barrier 
and $\Sigma$ its width. 
The condensate is moved towards the barrier 
by adding an initial momentum $p_0$ in the axial direction: 
\beq 
f(z,0) \to f(z,0) \; e^{- i p_0 z/\hbar} \; .  
\eeq 
In the case of a non-interacting condensate, 
i.e. $a_s=0$, the total 
energy per particle is practically the start-up 
kinetic energy $E_0=p_0^2/(2m)$, and the transverse 
energy $\hbar \omega_{\bot}$ does not affect the dynamics.  

\par 
As shown in Fig. 3, where we plot the density profile of 
the condensate along the symmetry axis at different instants, 
a fraction of the condensate 
tunnels the barrier while the rest is reflected. 
As expected, one sees also the interference between the 
incident wave function and the reflected wave function. 
Fig. 3 shows that the "exact" axial density profile obtained 
with the 3D GPE and that of the 1D NPSE are always quite close, 
also during the impact and tunneling time. 
These results suggest that 1D NPSE is very adequate 
also in the description of tunneling phenomena. 
\par 
The cigar-shaped configuration of the condensate 
is useful to study topological objects, 
like bright and dark solitons.  
By using 1D NPSE one finds out solitonic 
solutions which generalize what 
has been previously found with 1D GPE \cite{3,12}. 
Dark solitons ($a_s>0$) of Bose condensed atoms 
have been experimentally observed \cite{13}, 
while bright solitons ($a_s<0$) are more elusive 
due to the collapse of the condensate with a large 
number of atoms. 
\par 
Let us first consider bright solitons. 
Starting from our 1D NPSE, setting $V(z)=0$, 
scaling $z$ in units of $a_{\bot}$ and $t$ in units 
of $\omega_{\bot}^{-1}$, with the position 
\beq 
f(z,t)=\Phi(z-vt) e^{iv(z-vt)} e^{i(v^2/2 - \mu)t} \; , 
\eeq 
we find 
$$
\left[ {d^2\over d\zeta^2} - 2 \gamma 
{\Phi^2\over \sqrt{1-2\gamma\Phi^2} } \right. 
$$ 
\beq
\left. 
+ {1\over 2}   
\left( {1\over \sqrt{1-2 \gamma \Phi^2} } 
+ \sqrt{1-2\gamma \Phi^2} \right) \right] \Phi 
= \mu \Phi \; . 
\eeq 
where $\zeta =z-vt$ and $\gamma=|a_s|N/a_{\bot}$. 
This is a Newtonian second-order differential 
equation and its constant of motion is given by  
\beq 
E={1\over 2}\left({d\Phi\over d\zeta} \right)^2 
+ \mu \Phi^2 -\Phi^2\sqrt{1-2\gamma\Phi^2} \; .
\eeq 
The boundary condition $\Phi\to 0$ for 
$\zeta \to \infty$ implies that $E=0$. 
Then, by quadratures, one obtains the bright-soliton 
solution written in implicit form 
$$ 
\zeta= {1\over \sqrt{2}} {1\over \sqrt{1-\mu}} \; 
arctg\left[ 
\sqrt{ \sqrt{1-2\gamma\Phi^2}-\mu \over 1-\mu } 
\right] 
$$
\beq
-{1\over \sqrt{2}} {1\over \sqrt{1+\mu}} \; 
arcth\left[ 
\sqrt{ \sqrt{1-2\gamma\Phi^2}-\mu \over 1+\mu } 
\right] \; . 
\eeq 
Moreover, by imposing the normalization condition 
one also finds  
\beq 
(1-\mu)^{3/2} - {3\over2} (1-\mu)^{1/2} + 
{3\over 2 \sqrt{2}} \gamma = 0 \; . 
\eeq 
The normalization relates the chemical potential $\mu$ 
to the coupling constant $\gamma$, 
while the velocity $v$ of the bright soliton 
remains arbitrary. 
In the weak-coupling limit ($\gamma\Phi^2<<1$), 
the normalization condition gives 
$\mu= 1 - \gamma^2/2$ and the bright-soliton solution reads 
\beq
\Phi(\zeta)=\sqrt{\gamma\over 2} \; sech\left[{\gamma}\zeta \right] \; .  
\eeq 
The above solution is the text-book bright soliton 
of the 1D non-linear cubic Schr\"odinger equation (1D GPE). 
As shown in Figure 4, for 
$\sqrt{2}/3 < \gamma < 2/3$ there are two 
values for the chemical potential $\mu$ but we have numerically 
verified that lower one corresponds to an unstable 
solitonic solution. For $\gamma>2/3$ there are no 
solitary-wave solutions due to the collapse 
of the condensate. 

\begin{figure}
\centerline{\psfig{file=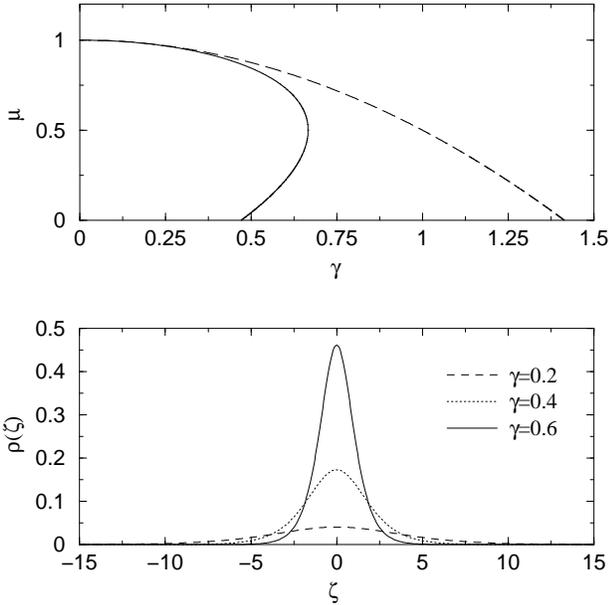,height=3.2in}}
\caption{On the top: chemical potential $\mu$ of the bright soliton 
as a function of the interaction strength 
$\gamma=|a_s|N/a_{\bot}$. 1D NPSE (full line) and 
1D GPE (dashed line). On the bottom: 
axial density profile $\rho(\zeta)=\Phi^2(\zeta)$ of the 
bright soliton of Bose condensed atoms for three 
values of the interaction strength $\gamma$ 
obtained with 1D NPSE.} 
\end{figure} 

\par
In the case of static dark solitons, the boundary conditions 
are: $\Phi(0)=0$ and 
$\Phi\to \Phi_0$ for $z\to\infty$. 
Starting from the 1D NPSE the analytical formula of 
the dark soliton is quite intricate because it 
involves elliptic integrals. Nevertheless, simple 
expressions can be found in the weak-coupling and 
in the strong-coupling limit. 
In the weak-coupling limits one finds  
\beq 
\Phi(z) = \Phi_0 \; 
th\left[ \Phi_0 \sqrt{2{\tilde \gamma}} z \right] \; , 
\eeq 
where $\Phi_0 = {\tilde \gamma}^{1/2}/\sqrt{2}$ 
and ${\tilde \gamma}=|a_s|\Delta N /a_{\bot}$, with 
$\Delta N$ the number of missing Bosons 
due to the hole in the condensate produced by the dark soliton. 
Eq. (20) is the familiar formula of stationary dark soliton 
for the 1D nonlinear Schr\"odinger equation (1D GPE), 
while in the strong-coupling limit one has 
\beq
|\Phi(z)| = {\Phi_0 \over 2} 
\left( 3 \; th^2\left[a|z|+b\right] -1 \right) \; , 
\eeq 
where $a=\sqrt{3}{\tilde \gamma}^{1/4}\sqrt{\Phi_0}/2^{3/4}$, 
$b=arcth[1/\sqrt{3}]$ and 
$\Phi_0={\tilde \gamma}^{1/6}/(\sqrt{2}(\sqrt{3}-2/3)^{2/3})$. 
\par
The study of the stability of our solitary-wave solutions is 
left to a future work. 3D numerical calculations \cite{14} 
suggest that Bose condensed dark solitons are stable for sufficiently 
small numbers of atoms or large transverse confinement. 
Moreover, a recent experiment \cite{15} has shown that 
dark solitons, created in a spherical Bose condensate,  
decay into vortex rings. 

\section{Effective 2D equation} 

In this section we derive 
the 2D reduction of the 3D GPE equation by using again a 
variational approach. In this case we take a trapping potential 
that is harmonic in the axial direction 
and generic in the transverse direction: 
$U({\bf r})=W(x,y)+{1\over 2}m\omega_z^2 z^2$. 
As trial wave-function we take the following 
\beq 
\psi({\bf r},t) = \phi(x,y,t) \; f(z,t;\eta(x,y,t)) 
\eeq
with 
\beq
f(z,t;\eta(x,y,t)) = 
{ e^{-z^2\over 2 \eta(x,y,t)^2} \over \pi^{1/4} \eta(x,y,t)^{1/2}} 
\; ,
\eeq 
where $\eta(x,y,t)$ is a variational function 
that describes the width of the condensate 
in the axial direction. 
We assume that the axial wavefunction $f$ is slowly 
varying along the transverse direction with respect 
to the axial direction, 
i.e. $\nabla^2 f \simeq {\partial^2 f\over \partial z^2}$.  
By inserting the trial wave-function in (2) and 
after spatial integration along $z$ variable, 
one finds the following Euler-Lagrange equations: 
$$ 
i\hbar {\partial \over \partial t}\phi= 
\left[ -{\hbar^2\over 2m}\nabla_{\bot}^2 
+ W + g N {\eta^{-1}\over \left(2\pi\right)^{1/2}} |\phi|^2 
\right. 
$$
\beq
\left. 
+ \left({\hbar^2 \over 2m}\eta^{-2} + {m\omega_z^2\over 2} \eta^2 
\right) \right] \phi  \; , 
\eeq  
\beq
{\hbar^2 \over 2m}\eta^{-3} - {1\over 2}m\omega_z^2 \eta 
+ g N {\eta^{-2}\over 2\left(2\pi\right)^{1/2}}|\phi|^2 = 0 \; . 
\eeq
Note that the second Euler-Lagrange equation 
can be written as 
$\eta^4 - 2\left(2\pi\right)^{1/2} a_z^4 a_s N |\phi|^2 \eta 
- a_z^4 = 0$, and, contrary to the case of 1D reduction, 
it does not have a elegant analytical solution 
but it can be easily solved \cite{16}. 
We name the 2D non-polynomial Schrodinger equation (25), 
with the condition (26), 2D NPSE.  
In Fig. 5 we compare the radial density profile $\rho(r)$ 
of the Bose condensate obtained by solving 
2D NPSE with the exact one obtained 
by solving the 3D GPE. For simplicity, we use 
a harmonic trapping 
potential $W(x,y)={1\over 2}m\omega_{\bot}^2(x^2+y^2)$ 
also in the transverse radial direction. 
Fig. 5 shows that 2D NPSE 
is quite reliable in describing the ground-state 
of disc-shaped Bose condensate. For example, 
the relative error in the calculation of the density 
at the orgin ranges from $0.7\%$ to $5\%$ 
by increasing $a_s/a_z$. 

\begin{figure}
\centerline{\psfig{file=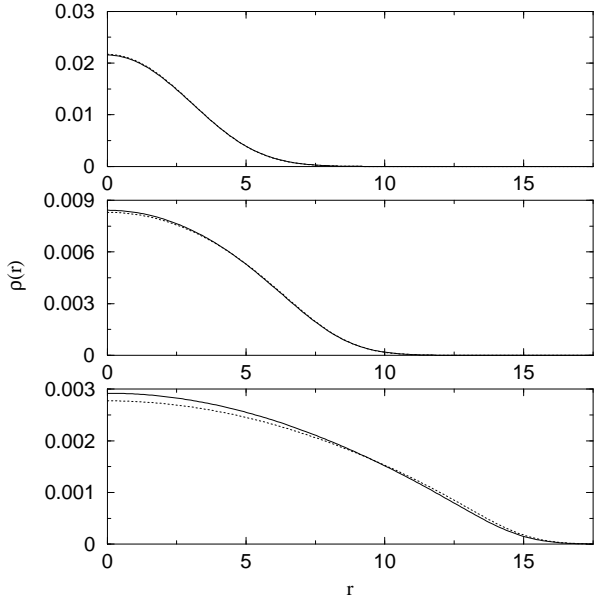,height=3.2in}}
\caption{Radial density profile 
$\rho(r)$ of the Bose condensate for a 
disc shaped trap. Number of Bosons: $N=10^4$ and 
trap anisotropy: $\omega_{\bot}/\omega_z=1/10$. 
3D GPE (solid line) and 2D NPSE (dotted line). 
From top to bottom: $a_s/a_z=10^{-4}$, 
$a_s/a_z=10^{-3}$, $a_s/a_z=10^{-2}$. 
Length $r=\sqrt{x^2+y^2}$ in units of $a_z$ and density 
in units $a_z^{-1}$.} 
\end{figure}

We can analyze the limits of weak and strong 
interaction of 2D NPSE. 
In the weakly-interacting case one finds 
$\eta =a_z$ and the equation of motion becomes 
\beq 
i\hbar {\partial \over \partial t}\phi= 
\left[ -{\hbar^2\over 2m} \nabla_{\bot}^2 
+ W + {g N \over \left(2\pi\right)^{1/2}a_z} |\phi|^2 
\right] \phi  \; , 
\eeq  
where the constant $\hbar \omega_z$ has been omitted because 
it does not affect the dynamics. This equation is a 
2D Gross-Pitaevskii equation. The nonlinear coefficient $g''$ 
of this 2D GPE can be thus obtained from the nonlinear coefficient 
$g$ of the 3D GPE by setting $g''=g/((2\pi)^{1/2} a_z)$. 
Eq. (27) describes a disc-shaped Bose condensate 
in a quasi-2D configuration. In fact, the weakly-interacting limit 
corresponds to a condensate with a chemical potential lower than 
$\hbar \omega_z$ \cite{8}. 
In the strongly-interacting case one has instead 
$\eta = \sqrt{2\pi^{1/3}} a_z^{4/3} a_s^{1/3} N^{1/3} 
|\phi|^{2/3}$, and the resulting nonlinear Sch\"odinger equation is 
\beq
i\hbar {\partial \over \partial t}\phi= 
\left[ -{\hbar^2\over 2m} \nabla_{\bot}^2 + W 
+ {3\over 4} 
{g N^{2/3} \over \pi^{2/3} a_z^{4/3} a_s^{1/3} } |\phi|^{4/3} 
\right] \phi  \; , 
\eeq
Note that in this limit, and in the stationary case, 
the kinetic term can be neglected (Thomas-Fermi approximation) 
and one finds the following analytical formula for 
the transverse density profile 
\beq 
|\phi(x,y)|^2 = {1\over 3\sqrt{3\pi}} 
{1\over (\hbar\omega_z)^{3/2} a_s a_z N} 
\left( \mu'' - W(x,y) \right)^{3/2} \; , 
\eeq 
where $\mu''$ is the chemical potential, fixed by the 
normalization condition. 
The same dependence is obtained starting 
from the Thomas-Fermi approximation of the stationary 3D GPE, 
i.e. neglecting the spatial derivatives in Eq. (1), 
and then integrating along $z$ variable. 
In this way one finds a formula that differs from 
the previous one only for the numerical factor which is 
$\sqrt{2}/(3\pi)$ instead of $1/(3\sqrt{3\pi})$.

\section*{Conclusions} 

We have found that the 1D 
non-polynomial nonlinear Schr\"odinger equation we have obtained 
by using a Gaussian variational ansatz from the 3D Gross-Pitaevskii 
action functional is quite reliable in describing the ground state 
and axial collective oscillations of 
cigar-shaped condensates. 
We have also tested the accuracy of the non-polynomial nonlinear 
equation in the case of a Bose condensate scattering and tunneling 
on a Gaussian barrier. The agreement with the 
results of the 3D Gross-Pitaevskii equation is very good 
for both ground-state and dynamics of the condensate. 
This equation will be useful for detailed numerical analysis 
of the {\sl dynamics} of cigar-shaped condensates, particularly 
when the local density may undergo sudden and large variations. 
In fact, the accurate mapping we have provided, allows to maintain 
a very good spatial resolution with modest 
computational effort even if the breakdown of the weak interaction condition 
during time evolution prevents the use of standard 1D GP equation. 
This case is often encountered in the study of reflection and tunneling 
events and in the propagation of solitary waves, 
which are currently being experimentally investigated. 
By using the 1D non-polynomial Schr\"odinger equation 
we have obtained analytical formulas for bright and 
dark solitons which generalize the ones known 
in the literature for the 1D non-linear cubic Schr\"odinger 
equation. Finally, we have deduced effective 2D equations describing 
the axial dynamics of disc-shaped condensates. 
From the 3D Gross-Pitaevskii action functional and using another 
Gaussian variational ansatz we have derived the effective 
2D nonlinear Schr\"odinger 
equation, which again has a non-polynomial structure. 
This structure simplifies in the weak and strong interacting 
regime.

\end{document}